\documentclass[aps]{revtex4}
\usepackage{amsmath}
\usepackage{bm}
\usepackage{array}
\usepackage{mathrsfs}
\usepackage{epstopdf}
\usepackage{amssymb}
\usepackage{amsfonts}
\usepackage{amssymb}
\usepackage{color}

\newcommand{\md}{\mathrm{d}}

\newcommand{\mF}{\mathcal{F}}
\newcommand{\mP}{\mathcal{P}}
\newcommand{\mV}{\mathcal{V}}
\newcommand{\mA}{\mathcal{A}}
\newcommand{\mS}{\mathcal{S}}

\begin{document}
\title{Spin Polarization Induced by Inhomogeneous Dynamical Condensate}
\author{Ziyue Wang$^{1}$}
\author{Pengfei Zhuang$^{1}$}
\affiliation{$^{1}$ Physics Department, Tsinghua University, Beijing 100084, China}
\date{\today}
\begin{abstract}
The role of dynamical chiral condensate in spin polarization is investigated in a kinetic theory framework. Transport equations for quark matter are derived in the mean-field approximation for the Nambu--Jona-Lasinio model. The dynamical condensate carries part of the energy momentum tensor (EMT) and the angular momentum tensor (AMT), the conservation of EMT and AMT can be proved from the kinetic equations as required by the symmetry. The transport equations of vector and axial-vector components are derived taking the spin decomposition as well as semi-classical expansion. Inhomogeneous mass introduces novel effect at $\mathcal{O}(\hbar)$, for an initially unpolarized system, spin polarization can be generated from the dynamical chiral condensate, even without the collision term. The stable spin distribution function is found to be robust, namely in the case with non-trivial dynamical mass, the spin polarization is still enslaved by the thermal vorticity, while the Killing condition can be loosened. 
\end{abstract}
\maketitle

\section{Introduction}
\label {s1}
The spin polarization effect in heavy ion collisions has attracted intense attention \cite{Liang:2004ph, Voloshin:2004ha, Betz:2007kg, Becattini:2007sr}. A large global angular momentum is produced in non-central heavy ion collisions and the spin of hadrons emitted is aligned with the direction of the global angular momentum \cite{STAR:2017ckg, Adam:2018ivw, Acharya:2019vpe}. The magnitude of the global polarization of $\Lambda$ baryons can be very well described by models based on relativistic hydrodynamics and assuming local thermodynamic equilibrium of the spin degrees of freedom \cite{Becattini:2013vja, Becattini:2015ska, Becattini:2016gvu, Karpenko:2016jyx, Pang:2016igs, Xie:2017upb}. However, discrepancy still exists for local polarization\cite{Niida:2018hfw}, which requires deeper understanding towards off-equilibrium phenomenon. Besides, the spin related anomalous transport phenomenon, such as chiral magnetic effect \cite{Kharzeev:2004ey, Fukushima:2008xe} and chiral vortical effect \cite{Neiman:2010zi} also call for the development of spin related transport theory and hydrodynamic theory. The chiral kinetic theory \cite{Son:2012bg, Son:2012wh, Son:2012zy, Stephanov:2012ki, Pu:2010as, Chen:2012ca, Hidaka:2016yjf, Huang:2018wdl, Liu:2018xip, Lin:2019ytz} is developed to describe the anomalous transport of massless fermions, and is further extended to the spin transport theory of massive fermions \cite{Hattori:2019ahi, Wang:2019moi, Gao:2019znl, Weickgenannt:2019dks, Liu:2020flb}. Recently, based on the Keldysh theory, it is extended from the free streaming scenario to the collisional effects \cite{Chen:2015gta, Yang:2020hri, Weickgenannt:2020aaf, Carignano:2019zsh, Li:2019qkf, Hou:2020mqp, Wang:2020pej}. In the researches above, the mass of fermion is taken to be zero (chiral limit) or a finite constant. However, the mass of light quarks comes mostly from chiral symmetry breaking, which is generated dynamically and is highly inhomogeneous in the heavy ion collision. The dynamically generated mass is well studies when investigating the phase transition in QGP, meanwhile, it is also found to appear in the transport equation of spin distribution \cite{Florkowski:1995ei, Huang:2020wrr}, but its relation to spin polarization has not been properly addressed. 

Particles get polarized in a rotation medium in a way which resembles the Barnett effect. In heavy ion collision, part of the initial orbital angular momentum (OAM) is transferred to the spin of particles. While the total angular momentum is conserved, the OAM and the spin are mutually convertible. From the microscopic view, such conversion takes place through collisions \cite{Zhang:2019xya}, which are described through collision terms in the kinetic theories. It has been found that spin get polarized by the 'nonlocal' collision terms in an initially unpolarized system \cite{Weickgenannt:2020aaf, Wang:2020pej}, which means the collision terms convert OAM with the spin. In our previous paper \cite{Wang:2020pej}, based of the Keldysh framework, collision terms are evaluated by adopting the Nambu--Jona-Lasinio (NJL) model and calculating the collisional self-energy. There, the collisions are considered perturbatively with the mean field self-energy ignored. It is well known that in equilibrium thermal field theory, the scattering above the mean field determines the width of the particle spectral function, while the mean field reflects the global properties of the system. Thus a question rises naturally, would the spin get polarized if one considers only the mean field self-energy in spin transport theory? And how would the dynamically generated mean field condensate changes the global equilibrium spin distribution? Qualitatively, if the mass of the particle changes by $\delta m$, the OAM would change by $\delta\vec L=(\delta m/m) \vec L$, such variance of OAM would be absorbed by the spin if the total angular momentum is conserved. Thus for considerable change of mass $(\delta m/m)$ around the chiral phase transition, the spin polarization would also be sizable. 

The global equilibrium spin distribution has been derived in several self-consistent ways, phenomenologically through the entropy production \cite{Becattini:2014yxa, Hattori:2019lfp, Fukushima:2020ucl}, and microscopicly through the collision terms in kinetic theory \cite{Weickgenannt:2020aaf, Wang:2020pej}. Spin polarization in global equilibrium is robust, namely from all the aforementioned methods, it is found to be enslaved by thermal vorticity $\varpi_{\mu\nu}=\partial_{[\mu} \beta_{\nu]}/2$, with $\beta^\mu=u^\mu/T$ satisfying the Killing condition $\partial^{\mu}\beta^{\nu}+\partial^{\nu}\beta^{\mu}=0$. In a realistic system, the nonzero thermal vorticity indicates the inhomogeneous distribution of temperature, and hence the inhomogeneous dynamical condensate. In this work, we take the inhomogeneous condensate into consideration, and try to find stable solution to the condensate modified collisionless spin transport equation. The global equilibrium spin distribution turns out to still be a stable solution, while the Killing condition is loosened. 

The paper is organized as follows: In Section \ref{s2}, we briefly review the Wigner-function approach and derive the kinetic equations after taking the mean field approximation. The classical and first order transport equations of vector and axial-vector components are derived by taking the semi-classical expansion.  In Section \ref{s3}, we derive the conservation of EMT and AMT from the kinetic equation. The stable solution of spin distribution function is discussed in Section \ref{s4}. Eventually, we make concluding remarks and outlook in Section \ref{s5}. For references, we present most of the details of computations and critical steps for derivations in the Appendix.

\section{Constraint and Transport Equation}
\label {s2}
In this section, we review the basic steps of deriving the transport equations with dynamical mass, and then take the semiclassical expansion to derive the classical and first order transport equations. 

Nambu--Jona-Lasinio (NJL) model at quark level describes well the chiral symmetry breaking in vacuum and its restoration at finite temperature and baryon density~\cite{Nambu:1961tp}. To investigate the dynamical chiral condensate in the quark transport theory, we adopt the one-flavor NJL model defined by,
\begin{eqnarray}
\mathcal L = \bar\psi\left(i\gamma^\mu \partial_\mu-m_0\right)\psi+G\left[(\bar\psi\psi)^2+(\bar\psi i\gamma_5\psi)^2\right], 
\end{eqnarray}
where $\psi$ is the one flavor quark field, $m_0$ is the quark current mass, and $G$ is the coupling constant. For simplicity, color and flavor degrees of freedom is neglected. Introducing the auxiliary field~\cite{Klevansky:1992qe} $\hat\sigma=-2G\bar\psi\psi$ and $\hat\pi=-2G\bar\psi i\gamma^5\psi$, the Lagrangian becomes, 
\begin{eqnarray}
\label{auxiL}
\mathcal{L}=\bar{\psi}\Big(\frac{1}{2}i\gamma^\mu\overleftrightarrow{\partial}_\mu-m_0-\hat{\sigma}-\hat{\pi}i\gamma^5\Big)\psi-\frac{\hat{\sigma}^2+\hat{\pi}^2}{4G}.
\end{eqnarray}
Under the mean field approximation, the Dirac equation of quark can be derived as $ 
[i\gamma^\mu \partial_\mu-m_0-\sigma(x)-i\gamma^5\pi(x)]\psi(x)=0$, where $\sigma(x)=\langle\hat\sigma\rangle$ and $\pi(x)=\langle\hat\pi\rangle$ are the mean field value of the auxiliary fields. In a typical situation, for a parity invariant ensemble or a parity invariant ground state, one would have $\langle\hat\pi\rangle=0$. However in a more general situation, especially for an out-of-equilibrium system, we can not exclude the $\langle\hat\pi\rangle\neq0$ from the very beginning. In order to study the transport phenomenon of spin, the covariant quark Wigner function $\mathcal{W}(x,p)$ is adopted, which is the ensemble average of the Wigner operator, 
\begin{eqnarray}
\mathcal{W}(x,p) &=& \int \md^4 y e^{-ipy}\Big\langle\psi(x+{y\over 2})\bar\psi(x-{y\over 2})\Big\rangle.
\end{eqnarray}
Calculating the first-order derivatives of the covariant density matrix in the Wigner function with respect to $x$ and $y$ and using the Dirac equation, one obtains the covariant kinetic equation,
\begin{eqnarray}
\label{kineticW}
[\gamma^\mu K_\mu+\gamma^5 K_5-M]\mathcal{W}(x,p) &=&0,
\end{eqnarray}
where $K_\mu=p_\mu+\frac{i\hbar}{2}\partial_\mu$, $K_5=\Pi_5+iD_5$ and $M=M_1+iM_2$. The operators are defined as 
\begin{eqnarray}
&&D_5=\cos(\hbar\Delta)\pi(x),\nonumber\\
&&M_1= m_0-\cos(\hbar\Delta)\sigma(x),\nonumber\\
&&\Pi_5= \sin(\hbar\Delta)\pi(x),\nonumber\\
&&M_2= \sin(\hbar\Delta)\sigma(x).
\end{eqnarray}
The operator $\Delta$ is defined as $\Delta=\frac{1}{2}\partial_\mu\partial_p^\mu$, with the derivative of coordinate acts only on the condensate, and the momentum derivative acts only on the Wigner function. Different Dirac components of the Wigner function have different physical meanings. Performing the spin decomposition of the Wigner function, one get various components as, 
\begin{eqnarray}
\label{decomposition}
\mathcal{W}(x,p) =\frac{1}{4}\left(\mathcal{F}+i\gamma_5 \mathcal{P}+\gamma_\mu \mathcal{V}^\mu+\gamma_\mu \gamma_5\mathcal{A}^\mu+\frac{\sigma_{\mu\nu}}{2}\mathcal{S}^{\mu\nu}\right).
\end{eqnarray}
In this self-consistent treatment of dynamically generated condensate, the mean field value of $\sigma(x)$ and $\pi(x)$ can be expressed in terms of the components of Wigner function as 
\begin{eqnarray}
\sigma(x)&=&-2G\int_p\text{Tr}\mathcal{W}(x,p)=-2G\int_p\mF(x,p),\nonumber\\
\pi(x)&=&-2G\int_p\text{Tr}\,i\gamma^5\mathcal{W}(x,p)=2G\int_p\mP(x,p).
\end{eqnarray}
Different from the equilibrium gap equation, the above expression for $\sigma(x)$ and $\pi(x)$ are valid even in off-equilibrium system. The inhomogeneous property of the $\sigma(x)$ and $\pi(x)$ directly comes from that of the Wigner function. From the kinetic equation of the Wigner function (\ref{kineticW}) as well as the spin decomposition, the transport and constraint equations for the spin components are
\begin{eqnarray}
\label{mastereq1}
p^\mu \mV_\mu-D_5\mP&=&M_1\mF,\nonumber\\
\hbar \partial^\mu \mA_\mu+2D_5\mF&=&2M_1\mP,\nonumber\\
p_\mu \mF-\frac{\hbar}{2}  \partial^\nu \mS_{\nu\mu}-\Pi_5 \mA_\mu&=&M_1\mV_\mu,\nonumber\\
-\hbar  \partial_\mu \mP+\epsilon_{\mu\nu\sigma\rho}p^\nu \mS^{\sigma\rho}-2\Pi_5\mV_\mu&=&2M_1 \mA_\mu,\nonumber\\
\frac{\hbar}{2} \partial_{[\mu} \mV_{\nu]}+\epsilon_{\mu\nu\sigma\rho}p^\sigma \mA^\rho-\frac{1}{2}\epsilon_{\mu\nu\sigma\rho}D_5\mS^{\sigma\rho}&=&M_1\mS_{\mu\nu},
\end{eqnarray}
as well as 
\begin{eqnarray}
\label{mastereq2}
\hbar  \partial^\mu \mV_\mu+2\Pi_5\mP&=&2M_2\mF,\nonumber\\
p^\mu \mA_\mu+\Pi_5\mF&=&-M_2\mP,\nonumber\\
\frac{\hbar}{2} \partial_\mu \mF+p^\nu \mS_{\nu\mu}-D_5\mA_\mu&=&M_2\mV_\mu,\nonumber\\
p_\mu \mP-D_5\mV_\mu+\frac{\hbar}{4}\epsilon_{\mu\nu\sigma\rho} \partial^\nu \mS^{\sigma\rho}&=&M_2\mA_\mu,\nonumber\\
p_{[\mu} \mV_{\nu]}-\frac{\hbar}{2}\epsilon_{\mu\nu\sigma\rho} \partial^\sigma \mA^\rho-\frac{1}{2}\epsilon_{\mu\nu\sigma\rho} \Pi_5\mS^{\sigma\rho}&=&M_2\mS_{\mu\nu}.
\end{eqnarray}
These equations has also been derived in previous researches with covariant Wigner function~\cite{Florkowski:1995ei} and equal-time Wigner function~\cite{Zhuang:1995jb}. 

The coupled constraint and transport equations Eq.(\ref{mastereq1}) and Eq.(\ref{mastereq2}) can not be solved for a general system. Thus in the following, we take the semi-classical expansion to consider the classical and quantum effect in the quark transport theory separately. Each component of the Wigner function (\ref{decomposition}) and the operators (\ref{kineticW}) are expanded by $\hbar$ and so as the constraint and transport equations Eq.(\ref{mastereq1}) and Eq.(\ref{mastereq2}). The components $\mathcal{V}$ and $\mathcal{A}$ give rise to the vector-charge and axial-charge currents. The 16 components given by the spin decomposition are not independent, takes the vector and axial vector components as independent degrees of freedom, the scalar component $\mathcal{S}$, pseudo-scalar component $\mathcal{P}$ and tensor component $\mathcal{S}_{\mu\nu}$ can be expressed in terms of them order by order in semiclassical expansion.

Take the classical limit of Eq.(\ref{mastereq1}) and Eq.(\ref{mastereq2}), the pion condensate is vanishing at $\mathcal{O}(\hbar^0)$ that $\pi^{(0)}=0$. The redundant degrees of freedom could be eliminated with respect to $\mV_\mu^{(0)}$ and $\mA_\mu^{(0)}$ by 
\begin{eqnarray}
&&\mF^{(0)}=\frac{p^\mu}{m} \mV^{(0)}_\mu,\nonumber\\
&&\mP^{(0)}=0,\nonumber\\
&&\mS_{\mu\nu}^{(0)}=\frac{1}{m}\epsilon_{\mu\nu\sigma\rho}p^\sigma \mA^{(0)\rho},
\end{eqnarray}
where $m=m_0-\sigma^{(0)}(x)$ is the quark mass in classical limit, with $m_0$ the current mass in the Lagrangian, and $\sigma^{(0)}(x)$ the dynamically generated chiral condensate. In the classical limit, $(p^2-m^2)\mV_\mu^{(0)}=0$ and $(p^2-m^2)\mA_\mu^{(0)}=0$ indicate that $\mV_\mu^{(0)}$ and $\mA_\mu^{(0)}$ are on-shell, with the mass-shell given by $m(x)=m_0-\sigma^{(0)}(x)$. Thus the classical components can be solved as $\mV_\mu^{(0)}=p_\mu f^{(0)}(x,p)\delta(p^2-m^2)$ and $\mA_\mu^{(0)}= f^{(0)}_{A\mu}(x,p)\delta(p^2-m^2)$. Each of $\mathcal{V}_\mu$ and $\mathcal{A}_\mu$ contains four components, which are not all independent. $p_\mu\mathcal{A}^{(0)\mu}=0$ and $p_{[\mu}\mathcal{V}^{(0)}_{\nu]}=0$ indicates that $\mathcal{A}^{(0)}_\mu$ has three independent components, while $\mathcal{V}^{(0)}_{\mu}$ has only one independent component. The transport equation of $\mV_\mu^{(0)}$ and $\mA_\mu^{(0)}$ is derived from the 4th and 5th equations of Eq.(\ref{mastereq2}) at $\mathcal{O}(\hbar)$-order, 
\begin{eqnarray}
\label{0th_transport_V}
p^\nu\partial_{\nu} \mV_{\mu}^{(0)}\,-\frac{(\partial_\mu m)}{m}p^\nu\mV_\nu^{(0)}\,\,+m(\partial^\nu m)(\partial_\nu^p\mV_\mu^{(0)})=0,\;\;\;\\
\label{0th_transport_A}
p^\nu\partial_{\nu} \mA^{(0)}_\mu+\frac{(\partial^\nu m)}{m}p_{[\mu}\mA^{(0)}_{\nu]}+m(\partial^\nu m)(\partial_\nu^p\mA_\mu^{(0)})=0.\;\;\;
\end{eqnarray}
These are classical Vlasov equation of number density and spin density in the covariant form, with the mass satisfying the gap equation at zeroth order given by 
\begin{eqnarray}
\label{classical_gap}
m\Big(1+2G\int_pf^{(0)}(x,p)\delta(p^2-m^2)\Big)=m_0.
\end{eqnarray}
It is worth noticing that the derivative on $\mV_{\mu}^{(0)}$ also acts on the $\delta$-function, that is to say $\partial_{\nu} \mV_{\mu}^{(0)}=p_\mu (\partial_{\nu} f^{(0)})\delta(p^2-m^2)-2m(\partial_{\nu} m)p_\mu f^{(0)}(x,p)\delta'(p^2-m^2)$, and similar for $\partial_{\nu}^p \mV_{\mu}^{(0)}$. Considering the on-shell condition at classical level, and that the equal-time number density and spin density are related to covariant components by $f_0^\pm(x,\vec p)=\int dp_0 \mV_0(x,p)\delta(p_0\mp E_p)$ and $\vec{g}_0(x,\vec p)=-\int dp_0 \mA_i(x,p)\delta(p_0\mp E_p)$, then one can recover the familiar formulae of Vlasov equation~\cite{Zhuang:1995jb}
\begin{eqnarray}
\label{f0g0}
&&\Big(\partial_{t} \pm {\vec p\over E_p}\cdot{\nabla}\mp {{\bf \nabla}m^2\cdot{\bf \nabla}_p\over 2E_p}\Big)f_0^{(0)\pm} = 0,\\
&&\Big(\partial_{t} \pm {\vec p\over E_p}\cdot{\nabla}\mp {{\bf \nabla}m^2\cdot{\bf \nabla}_p\over 2E_p}\Big){\vec g}_{0}^{(0)\pm} \nonumber\\
&&=- {1\over 2m^2E_p^2}\left(\partial_tm^2{\vec p}\pm E_p{\bf \nabla}m^2\right)\times\big({\vec p}\times{\vec g}_{0}^{(0)\pm}\big).
\end{eqnarray}

The investigation of spin transport phenomenon requires the transport equations of the first order components, we thus expand the constraint and transport equations Eq.(\ref{mastereq1}) and Eq.(\ref{mastereq2}) to the first order of $\hbar$. The independent components are still taken to be $\mV_\mu^{(1)}$ and $\mA_\mu^{(1)}$, the scalar, pseudo-scalar, and the tensor components can be expressed in terms of vector as well as axial vectors as
\begin{eqnarray}
&&\mF^{(1)}=\frac{p^\mu}{m} \mV_\mu^{(1)}+\frac{\sigma^{(1)}}{m}\mF^{(0)},\nonumber\\
&&\mP^{(1)}=\frac{1}{2m}\nabla^\mu \mA_\mu^{(0)}+\frac{\pi^{(1)}}{m}\mF^{(0)},\nonumber\\
&&\mS_{\mu\nu}^{(1)}=\frac{1}{2m}\nabla_{[\mu} \mV_{\nu]}^{(0)}+\frac{1}{m}\epsilon_{\mu\nu\sigma\rho}p^\sigma \mA^{(1)\rho}\nonumber\\
&&\qquad\quad +\frac{\pi^{(1)}}{m^2}p_{[\mu}\mA^{(0)}_{\nu]}+\frac{\sigma^{(1)}}{m^2}\epsilon_{\mu\nu\sigma\rho}p^\sigma\mA^{(0)\rho}.
\end{eqnarray}
The mass shell condition at first order is modified by the condensate $\sigma^{(1)}$, 
\begin{eqnarray}
(p^\nu p_\nu-m^2)\mV_\mu^{(1)}&=&-2m\sigma^{(1)}\mV_\mu^{(0)},\nonumber\\
(p^\nu p_\nu -m^2)\mA_\mu^{(1)}&=&-2m\sigma^{(1)}\mA_\mu^{(0)},
\end{eqnarray}
consider the classical solution of $\mV_\mu^{(1)}$ and $\mA_\mu^{(1)}$, and use the property $x\delta'(x)=-\delta(x)$, the solutions for $\mV_\nu^{(1)}$ and $\mA_\nu^{(1)}$ are
\begin{eqnarray}
\mV_\mu^{(1)}&=&p_\mu f^{(1)}\delta(\xi)+2\frac{\sigma^{(1)}}{m}p_\mu f^{(0)}\delta'(\xi), \\
\label{onshellA1}
\mA_\mu^{(1)}&=& f_{A\mu}^{(1)}\delta(\xi)+2\frac{\sigma^{(1)}}{m}f_{A\mu}^{(0)}\delta'(\xi), 
\end{eqnarray}
with $\xi$ the shorthand notation for $p^2-m^2$. Thus the mass shell of $\mV_\mu=\mV_\mu^{(0)}+\hbar\mV_\mu^{(1)}$ is modified to be $[p^\nu p_\nu-(m-\hbar\sigma^{(1)})^2] (\mV_\mu^{(0)}+\hbar\mV_\mu^{(1)})=0$. $\mV_\mu^{(1)}$ is off-shell, but $\mathcal{V}_\mu=\mathcal{V}_\mu^{(0)}+\hbar\mathcal{V}_\mu^{(1)}$ are still on the modified mass-shell, and the same mass-shell condition for $\mA_\mu$. The transport equation of $\mV_\mu^{(1)}$ and $\mA_\mu^{(1)}$ is derived from the 4th and 5th equations of Eq.(\ref{mastereq2}) at $\mathcal{O}(\hbar^2)$-order, the transport equation of $\mV_\mu^{(1)}$ is 
\begin{eqnarray}
\label{1st_transport_V}
&&p^\nu \nabla_{\nu} \mV_{\mu}^{(1)}
-\frac{(\nabla_\mu m)}{m}p^\nu\mV_\nu^{(1)}
+m(\nabla^\nu m)(\nabla^\nu_p\mV_\mu^{(1)})
\nonumber\\
&&=
-(\nabla_\mu\frac{\sigma^{(1)}}{m})p^\nu\mV^{(0)}_\nu
+(\nabla_\nu m\sigma^{(1)})(\nabla^\nu_p\mV_\mu^{(0)})\\
 &&
 -\frac{\pi^{(1)}}{m}\epsilon_{\mu\nu\sigma\rho}p^\nu \nabla^{\sigma} \mV^{(0)\rho} 
+\frac{1}{2m}\epsilon_{\mu\nu\sigma\rho}p^\nu(\nabla^\lambda\nabla^\sigma m)\nabla^\lambda_p\mA^{(0)\rho},\nonumber
\end{eqnarray}
as well as the transport equation of $\mA_\mu^{(1)}$, 
\begin{eqnarray}
\label{1st_transport_A}
&&p_\nu \nabla^\nu \mA^{(1)}_{\mu}
+\frac{(\nabla^\nu m)}{m}p_{[\mu} \mA^{(1)}_{\nu]}
+m(\nabla^\nu m)(\nabla_\nu^p\mA_\mu^{(1)})
\nonumber\\
&&=
(\nabla_\nu\frac{\sigma^{(1)}}{m})p_{[\mu}\mA^{(0)}_{\nu]}
+(\nabla^\nu \sigma^{(1)}m)(\nabla_\nu^p\mA_\mu^{(0)})\\
&&
-m\epsilon_{\mu\nu\sigma\rho}\Big(\nabla^\nu \frac{\pi^{(1)}}{m^2}p^{\sigma}\mA^{(0)\rho}\Big)
+\frac{1}{2m}\epsilon_{\mu\nu\sigma\rho}(\nabla^\nu m)(\nabla^{\sigma} \mV^{(0)\rho}),\nonumber
\end{eqnarray}
with the condensate at first order determined by, 
\begin{eqnarray}
\label{1st_condensate}
\sigma^{(1)}&=&-2Gm\left(\int_p f^{(1)}\delta(\xi)\right)\nonumber\\
&&\times\left[1+4G\int_p\frac{p^\mu p_\mu}{m^2} f^{(0)}\delta'(\xi)+2G\int_pf^{(0)}\delta(\xi)\right]^{-1},\nonumber\\
\pi^{(1)}&=&\frac{2G}{2m}\nabla^\mu \int_pf_{A\mu}^{(0)}\delta(\xi)+2G\pi^{(1)}\int_pf^{(0)}(x,p)\delta(\xi).
\end{eqnarray}
Eq.(\ref{1st_transport_V}, \ref{1st_transport_A}) together with Eq.(\ref{1st_condensate}) determines the transport phenomenon of the vector and axial-vector charge at first order. The LHS has the same structure as the classical transport equations (\ref{0th_transport_V}, \ref{0th_transport_A}), while the RHS contains nontrivial terms leading to coupling transport of vector and axial-vector component, namely in the transport equation of $\mV_\mu^{(1)}$, $\mA_\mu^{(0)}$ is involved, and vice versa. At first order, the coupling arises from the inhomogeneous dynamically generated chiral condensate. One would expect that, more complicated coupling terms would appear in transport equations at higher order of $\hbar$, but here we restrict ourself to $\mathcal{O}(\hbar)$-order, where the nontrivial effect already appears.

Consider an initially unpolarized system with $\mA_\mu=0$, the zeroth order transport equation (\ref{0th_transport_A}) would keep the system unpolarized $\mA_\mu^{(0)}=0$. While the first order transport equation (\ref{1st_transport_A}) generates nonzero spin polarization $\mA_\mu^{(1)}\neq0$ due to the last term in the RHS $\frac{1}{2m}\epsilon_{\mu\nu\sigma\rho}(\nabla^\nu m)(\nabla^{\sigma} \mV^{(0)\rho})$. In a realistic system, the dynamically generated mass $m(x)=m_0-\sigma(x)$ is inhomogeneous, $\mV^{(0)\rho}$ reflecting the number density is also nonzero, thus this coupling term would be nonzero for any general case. During the off-equilibrium evolution, this term would generate nonzero $\mA_\mu^{(1)}$, hence lead to spin polarization in the system. In a previous paper \cite{Wang:2020pej}, we have proved that for an initially unpolarized system, spin polarization can be generated from collision terms at $\mathcal{O}(\hbar)$. Specifically, we adopted a four fermion interaction, which is the $\sigma$-channel in the NJL model, the collision terms were calculated in the case of vanishing meanfield self-energy. Since the same interaction also brings about chiral condensate when the temperature is not too high and the density is not too large. A natural question is that, in an off-equilibrium system, would the condensate play a similar role in producing spin polarization? Here, the collisionless first order transport equation (\ref{1st_transport_A}) indeed shows that the spin polarization could also be generated from the dynamical chiral condensate. 

In the discussion of collision terms\cite{Wang:2020pej}, another interesting problem is to obtain the equilibrium solution to the transport equation. The detailed balance principle requires the cancellation of the gain term and loss term cancel at local equilibrium, the equilibrium spin distribution can be derived accordingly, and is found to be robust. A similar question here is that, whether the spin polarization generated from the chiral condensate would survive until the equilibrium? In order to answer this question, we try to find the stable solution to the first order transport equation (\ref{1st_transport_A}), before that, it is necessary to check the conservation of energy momentum tensor (EMT) and angular momentum tensor (AMT) from the kinetic equation. 

\section{Conservation of EMT \& AMT}
\label {s3}
The energy momentum tensor can be obtained through the Lagrangian by
\begin{eqnarray}
T^{\mu\nu}&=&\frac{\partial\mathcal{L}}{\partial(\partial_\mu\psi)}\partial^\nu\psi+\partial^\nu\psi^\dag\frac{\partial\mathcal{L}}{\partial(\partial_\mu\psi^\dag)}-g^{\mu\nu}\mathcal{L}, 
\end{eqnarray}
since we are going to consider the condensation at mean field level, we take the Lagrangian with the auxiliary field (\ref{auxiL}). Then, the ensemble average of the energy momentum tensor involves contribution from the fermion field as well as the condensation,  
\begin{eqnarray}
\langle T^{\mu\nu}\rangle
&=&\frac{i }{2}\langle\bar{\psi}\gamma^\mu\overleftrightarrow{\partial}^\nu\psi\rangle
-g^{\mu\nu}\Big\langle\bar{\psi}(\frac{1}{2}i\gamma^\mu\overleftrightarrow{\partial}_\mu-m_0)\psi\Big\rangle\nonumber\\
&&+g^{\mu\nu}\big\langle\hat{\sigma}\bar{\psi}\psi-\hat{\pi}\bar{\psi}i\gamma^5\psi\big\rangle+g^{\mu\nu}\frac{\langle\hat{\sigma}\rangle^2+\langle\hat{\pi}\rangle^2}{4G}.\;
\end{eqnarray}
Terms in the first line comes from the free fermion part, while for the second line, the ensemble average is evaluated by considering that $
\langle\sigma(x)\bar{\psi}(x)\psi(x)\rangle
=-\int d^4 p\cos(\hbar \Delta)\sigma(x) \mF(x,p)$, for the proof see (\ref{proof1}) in Appendix.$\ref{appen1}$, and similar for $\langle\pi(x)\bar{\psi}(x)i\gamma_5\psi(x)\rangle$. We have also used the first equation in Eq.(\ref{mastereq1}) that $p^\mu \mV_\mu-m_0\mF=\cos(\hbar\Delta)\pi(x)\mP-\cos(\hbar\Delta)\sigma(x)\mF$. The ensemble average of the canonical energy momentum tensor is evaluated to be
\begin{eqnarray}
\label{EMT}
\langle T^{\mu\nu}\rangle
&=&\int d^4 p p^\nu \mV^\mu+g^{\mu\nu}\frac{{\sigma(x)}^2+{\pi(x)}^2}{4G}.
\end{eqnarray}
The first term is the fermion part, which is the same as that of free fermion system, while the second term comes from the mean field of condensation, which does not appear in free fermion system. The conservation of energy momentum tensor requires that
\begin{eqnarray}
\label{EMT_conservation1}
\partial_\mu \langle T^{\mu\nu}\rangle
=\int d^4 p p^\nu \partial_\mu \mV^\mu+\frac{{\sigma}\partial^\nu{\sigma} +{\pi}\partial^\nu{\sigma}}{2G}=0.
\end{eqnarray}
Since the collision terms are not included, the kinetic equation (\ref{mastereq1}) and (\ref{mastereq2}) should also reproduce the conservation of energy momentum tensor. Multiply the first equation of (\ref{mastereq2}) by $p^\nu$ and then take integral over four momentum, we obtain, 
\begin{eqnarray}
\label{EMT_conservation2}
\int d^4 pp^\nu \partial_\mu \mV^\mu+\frac{2}{\hbar}\int d^4 pp^\nu \sin(\hbar\Delta) \big[\pi\,\mP-\sigma\,\mF\big]=0, \;
\end{eqnarray}
the condensate $\pi$ and $\sigma$ have coordinate dependence, the coordinate derivative in $\Delta$ acts only on the condensate, while the momentum derivative acts only on the components of Wigner function. Considering that the condensate can be obtained from the scalar and axial-scalar component $\sigma=-2G\int_p\mF$ and $\pi(x)=2G\int_p\mP$, and that $
\int d^4p p^\nu\sin(\hbar\Delta)\sigma(x) W(x,p)=\frac{1}{2}\sigma(x)\int  d^4p\partial^\nu W(x,p)$ (for the proof, please see (\ref{proof2}) in Appendix.$\ref{appen1}$), one can prove that Eq.(\ref{EMT_conservation1}) and Eq.(\ref{EMT_conservation2}) are exactly the same, namely
\begin{eqnarray}
\partial_\mu\langle T^{\mu\nu}\rangle
&=&\int d^4 p p^\nu \partial_\mu \mV^\mu+\frac{{\sigma}\partial^\nu{\sigma} +{\pi}\partial^\nu{\sigma}}{2G}\nonumber\\
&=&\int d^4 p\Big[ p^\nu \partial_\mu\mV^\mu-\sigma(x)\partial^\nu\mF +\pi(x)\partial^\nu\mP\Big]\nonumber\\
&=&\int d^4 p\Big\{p^\nu \partial_\mu \mV^\mu+\frac{2}{\hbar}p^\nu \sin(\hbar\Delta) \big[\pi\,\mP-\sigma\,\mF\big]\Big\}\nonumber\\
&=&\int d^4 p p^\nu[ \partial_\mu\mV^\mu-2M_2\mF+2\Pi_5\mP]\nonumber\\
&=&0.
\end{eqnarray}
We would like to emphasize here that in the current scenario with the dynamically generated condensate, $\int d^4 p p^\nu \partial_\mu \mV^\mu$ is only a part of the energy momentum tensor, and is not conserved. During the transportation, energy converts between fermion field and the condensate, while the total amount is conserved. 

The canonical total angular momentum tensor is $J^{\lambda,\mu\nu}=L^{\lambda,\mu\nu}+S^{\lambda,\mu\nu}$, where $L^{\lambda,\mu\nu}=x^\mu T^{\lambda\nu}-x^\nu T^{\lambda\mu}$ is the canonical orbital angular momentum (OAM) and $S^{\lambda,\mu\nu}$ is the canonical spin tensor, which is defined by 
\begin{eqnarray}
S^{\lambda,\mu\nu}=-\frac{i}{2}\frac{\partial \mathcal{L}}{\partial(\partial_\lambda\psi)}\sigma^{\mu\nu}\psi+\frac{i}{2}\bar{\psi}\sigma^{\mu\nu}\frac{\partial \mathcal{L}}{\partial(\partial_\lambda\bar\psi)}.
\end{eqnarray}
The ensemble average of the spin tensor can be expressed in terms of the Wigner function by $\langle S^{\lambda,\mu\nu}\rangle=\frac{\hbar}{2}\int d^4 p\epsilon^{\lambda\mu\nu\alpha}\mA_\alpha$. The condensate contributes to the OAM and spin only by affecting the mass of the fermion field. Using Noether’s theorem, the invariance of the action under the rotation transformations yields the conservation of
the canonical total angular momentum tensor $J^{\lambda,\mu\nu}$, which requires,
\begin{eqnarray}
\partial_\lambda J^{\lambda,\mu\nu}= T^{[\mu\nu]}+\partial_\lambda S^{\lambda,\mu\nu}=0, 
\end{eqnarray}
since the condensation part in energy momentum tensor is symmetric, the antisymmetric part $T^{[\mu\nu]}$ contains only the fermion part. The conservation of total angular momentum can also be derived from the kinetic equation. Taking momentum integral over the last equation in (\ref{mastereq2}), we obtain,
\begin{eqnarray}
\label{AMT_conservation}
T^{[\mu\nu]}+\partial_\lambda S^{\lambda,\mu\nu}
&=&-\frac{1}{2}\int d^4 p\epsilon^{\mu\nu\sigma\rho} \sin(\hbar\Delta)\pi(x)\mS_{\sigma\rho}\nonumber\\
&&-\int d^4 p\sin(\hbar\Delta)\sigma(x)\mS^{\mu\nu}\nonumber\\
&=&0,
\end{eqnarray}
notice that $S^{\lambda,\mu\nu}$ on the LHS is the spin tensor, while $\mS^{\mu\nu}$ on the RHS is the tensor component of the Wigner function. The condensation seems to break to AMT conservation, however one can show that RHS is exactly vanishing considering that $\int d^4p \sin(\hbar\Delta)\sigma(x) W(x,p)=0$ and the same for the term involving $\pi$-condensate, the proof is presented in (\ref{proof3}) in Appendix.$\ref{appen1}$. 

\section{stable solution}
\label {s4}
For off-equilibrium system, collisions brings the system to equilibrium and are crucial in the transport phenomenon. When the system reaches equilibrium, the collision term exactly vanishes with the gain and loss term cancel with each other. Meanwhile the free-streaming part of the transport equation also vanishes. Since the kinetic equations are derived in the collisionless limit, the classical and first order transport equations can be treated as the free-streaming part in local equilibrium when the collision term exactly vanishes. Thus the stable solution of the transport equation (\ref{0th_transport_V}, \ref{0th_transport_A}) and (\ref{1st_transport_V}, \ref{1st_transport_A}) would illustrate how the inhomogeneous condensate affects the equilibrium distribution. 

In order to simplify the problem, we also consider an initially unpolarized system with $\mA_\mu(t=0)=0$. Transport equation of $\mA_\mu^{(0)}$ indicates that $\mA_\mu^{(0)}$ remains zero during the evolution of the system, and thus is vanishing in the equilibrium state. Setting $\mA_\mu^{(0)}=0$ would further simplify the problem.  First, from the gap equation of $\pi$ condensate, 
\begin{eqnarray}
\pi^{(1)}(x)\left(1-2G\int_pf^{(0)}(x,p)\delta(\xi)\right)=0,
\end{eqnarray}
the pion condensate at $\mathcal{O}(\hbar)$ has only trivial solution $\pi^{(1)}=0$. With vanishing $\mA_\mu^{(0)}$, the first order transport equations also get simplified 
\begin{eqnarray}
\label{1st_transport_V1}
&&p^\nu \nabla_{\nu} \mV_{\mu}^{(1)}
-\frac{\nabla_\mu m}{m} p^\nu\mV_\nu^{(1)}
+m\nabla_\nu m\nabla^\nu_p\mV_\mu^{(1)}\nonumber\\
&=&
-(\nabla_\mu\frac{\sigma^{(1)}}{m})p^\nu\mV^{(0)}_\nu
+(\nabla_\nu m\sigma^{(1)})(\nabla^\nu_p\mV_\mu^{(0)}),\\
\label{1st_transport_A1}
&&p_\nu \nabla^\nu \mA^{(1)}_{\mu}
+\frac{(\nabla^\nu m)}{m}p_{[\mu} \mA^{(1)}_{\nu]}
+m(\nabla^\nu m)(\nabla_\nu^p\mA_\mu^{(1)})
\nonumber\\
&=&
\frac{1}{2m}\epsilon_{\mu\nu\sigma\rho}(\nabla^\nu m)(\nabla^{\sigma} \mV^{\rho(0)}).\qquad\qquad
\end{eqnarray}
It is worth noticing that transport equation (\ref{1st_transport_A1}) is numerically solvable, as in the RHS, the classical mass $m$ is determined by the classical gap equation (\ref{classical_gap}), which involves only the classical number density distribution determined by the classical transport (\ref{0th_transport_V}). In a numerical solution, one can first solve the classical transport (\ref{0th_transport_V}) together with the gap equation (\ref{classical_gap}) to determine the evolution of mass as well as the distribution function\cite{Wang:2020wwm}, then substitute into (\ref{1st_transport_A1}) to work out the evolution of spin density. Besides, from the on-shell relation of $\mA_\mu^{(1)}$ Eq.(\ref{onshellA1}), vanishing zeroth order component guarantees the on-shell condition of $\mA_\mu^{(1)}$, that $\mA_\mu^{(1)}=f_{A\mu}^{(1)}\delta(p^2-m^2)$, with $f_{A\mu}^{(1)}$ some unknown function in phase space. From the second equation in Eq.(\ref{mastereq2}) $p^\mu\mA_\mu^{(1)}=0$, $\mA_\mu^{(1)}$ has three degrees of freedom. Hence it can be parametrized as $\mA_\mu^{(1)}=m\theta_\mu f_A\delta(\xi)$, where $f_A$ is the axial distribution function, and $\theta_\mu$ is a unit vector, which is normalized with spacelike condition $\theta^\mu\theta_\mu=-1$ and $p^\mu\theta_\mu=0$. The spin tensor is then $\Sigma_S^{\mu\nu}=\frac{1}{2m}\epsilon^{\mu\nu\rho\sigma}\theta_\rho p_\sigma$. 

In order to construct a general ansatz for the local equilibrium distribution function, we start from the requirement that the distribution functions must depend only on the linear combination of the collisional conserved quantities \cite{Liu:2020flb}. Consider a simple case of vanishing baryon number density, the general ansatz is taken to be $f_{\pm}=f(g_\pm)$ with $g_\pm=p\cdot\beta+\hbar\Sigma_S^{\mu\nu}\omega_{\mu\nu}$ for massive fermion. The Lagrange multiplier $\beta^\mu$, $\omega_{\mu\nu}$ depend only on $x$. $f_{\pm}$ are the distribution function parallel and anti-parallel to the polarization direction $\theta_\mu$ respectively. The vector and axial charge distribution can then be obtained by $f_{V/A}=(f_{+}\pm f_{-})/2$. Thus in general, the classical vector charge distribution and the first order spin distribution function can be constructed as 
\begin{eqnarray}
\label{ansatz}
V^{(0)}_{\mu}
&=&p_\mu f(p\cdot\beta)\delta(p^2-m^2),\nonumber\\
A^{(1)}_{\mu}
&=&G\epsilon_{\rho\sigma\lambda\mu}p^\rho\omega^{\sigma\lambda} f_A(p\cdot\beta)\delta(p^2-m^2),
\end{eqnarray}
where the spin chemical potential $\omega^{\sigma\lambda}$ is some unknown antisymmetric tensor, and is a function of coordinate, but not momentum; $f$ and $f_A$ are some unknown functions of collisional invariant $\beta\cdot p$; G is a dimensionless normalization constant.

First, we need to check, how would the the free-streaming zeroth order equation (\ref{0th_transport_V}) sets constraint to Lagrange multiplier $\beta^\mu$. By substituting the above ansatz (\ref{ansatz}) in to the transport equation of $V^{(0)}_{\mu}$ Eq.(\ref{0th_transport_V}), 
\begin{eqnarray}
\Big[ p^\rho p^\nu\nabla_{\nu}\beta_\rho+m \beta^\nu\nabla_\nu m\Big] f'\delta(\xi)=0.
\end{eqnarray}
where $m(x)=m_0-\sigma^{(0)}$ is the classical mass determined by the classical gap equation (\ref{classical_gap}). The vanishing of free-streaming part indicates that terms in the square bracket vanishes under the constraint of delta function, 
\begin{eqnarray}
\label{requirement0}
p^\rho p^\nu(\nabla_{\nu}\beta_\rho+\nabla_{\rho}\beta_\nu)=-2m \beta^\mu\nabla_\mu m.
\end{eqnarray}
If mass is homogeneous in the system, the RHS is zero, and the Killing condition is retained $\nabla_{\nu}\beta_\rho+\nabla_\rho\beta_{\nu}=0$; if the gradient of mass is perpendicular to $\beta^\mu$, Killing condition also exist. However, for a general system, Killing condition no longer exist, but as long as (\ref{requirement0}) if fullfilled, the solution would be stable. Thus (\ref{requirement0}) is the constraint set to the Lagrange multiplier $\beta_\mu$.

Then we are going to find out the constraint set to the spin chemical potential. Substituting the ansatz $A^{(1)}_{\mu}$ (\ref{ansatz}) into the first order transport equation (\ref{1st_transport_A1}), and using (\ref{requirement0}), the first order transport equation requires that 
\begin{eqnarray}
\label{requirement1}
&&
G\epsilon_{\rho\sigma\lambda\mu}(p^\rho p^\nu\nabla_\nu\omega^{\sigma\lambda}) f_A\delta(\xi)\nonumber\\
&&+Gm(\nabla^\rho m) \epsilon_{\rho\sigma\lambda\mu}\omega^{\sigma\lambda} f_A\delta(\xi)
\nonumber\\
&&+G\frac{(\nabla^\nu m)}{m}p_{\mu} \epsilon_{\rho\sigma\lambda\nu}p^\rho\omega^{\sigma\lambda} f_A\delta(\xi)\\
&&-G\frac{(p_{\nu} \nabla^\nu m)}{m}\epsilon_{\rho\sigma\lambda\mu}p^\rho\omega^{\sigma\lambda} f_A\delta(\xi)\nonumber\\
&&-\frac{1}{2m}\epsilon_{\mu\nu\sigma\rho}p^\rho(\nabla^\nu m)(p_\lambda\nabla^{\sigma}\beta^\lambda) f'\delta(\xi)=0.\nonumber
\end{eqnarray}
Using the Schouten identity $p_{\mu}\epsilon_{\rho\sigma\lambda\nu}
=-p_{\rho}\epsilon_{\sigma\lambda\nu\mu}-p_{\sigma}\epsilon_{\lambda\nu\mu\rho}-p_{\lambda}\epsilon_{\nu\mu\rho\sigma}-p_{\nu}\epsilon_{\mu\rho\sigma\lambda}$ to rewrite the third term, Eq.(\ref{requirement1}) becomes, 
\begin{eqnarray}
\label{requirement2}
&&
2mG\epsilon_{\rho\sigma\lambda\mu}(p^\rho p^\nu\nabla_\nu\omega^{\sigma\lambda}) f_A\delta(\xi)\nonumber\\
&&
-\epsilon_{\mu\nu\sigma\rho}p^\rho(\nabla^\nu m)(p_\lambda\nabla^{\lambda}\beta^\sigma) f'\delta(\xi)\nonumber\\
&&
-4G\epsilon_{\mu\nu\sigma\rho}(\nabla^\nu m)p^\rho(p_{\lambda}\omega^{\sigma\lambda}) f_A\delta(\xi)\nonumber\\
&&
-\epsilon_{\mu\nu\sigma\rho}p^\rho(\nabla^\nu m)(p_\lambda\nabla^{[\sigma}\beta^{\lambda]}) f'\delta(\xi)
=0.
\end{eqnarray}
One solution that returns naturally to constant mass scenario is taking $f_A=f'$ and let $\omega^{\sigma\lambda}=\frac{1}{2}\nabla^{[\sigma}\beta^{\lambda]}$, with $G=-1/2$. With these choices, the last two terms in (\ref{requirement2}) cancel with each other, leaving
\begin{eqnarray}
\label{requirement3}
\epsilon_{\mu\nu\sigma\rho}p^\rho \Big[m\nabla^{\nu}p_\lambda\nabla^\lambda\beta^{\sigma}
-(\nabla^\nu m)(p_\lambda\nabla^{\lambda}\beta^\sigma)\Big] f'\delta(\xi)
=0.\;\;
\end{eqnarray}
The terms in the square bracket then should be vanishing, using the constraint set to $\beta_\mu$ (\ref{requirement0}), we obtain
\begin{eqnarray}
\label{requirement4}
\nabla^\mu(\beta^\nu\nabla_\nu m)=0.
\end{eqnarray}
This indicates that $\beta^\nu\nabla_\nu m$ must be independent of $x$.

Thus under the constraints set to $\beta_\mu$ (\ref{requirement0}) and (\ref{requirement4}), the spin polarization generated from the condensate is still enslaved by the thermal vorticity $\varpi^{\sigma\lambda}=\frac{1}{2}\nabla^{[\sigma}\beta^{\lambda]}$, with the stable solution takes the same form as in the case of constant mass,
\begin{eqnarray}
\label{stablesolution}
A^{(1)}_{\mu}
&=&\frac{1}{2}\epsilon_{\mu\rho\sigma\lambda}p^\rho\varpi^{\sigma\lambda}f'(p\cdot\beta)\delta(\xi).
\end{eqnarray}
With the dynamically generated mass, the vanishing of the free-streaming part no longer requires the Killing condition. The Killing condition is now loosened to be $[p^\rho p^\nu\nabla_{\nu}\beta_\rho+m \beta^\nu\nabla_\nu m] \delta(\xi)=0$ with $\beta^\nu\nabla_\nu m$ independent of $x$. A special case is that $\beta^\nu\nabla_\nu m$ is exactly zero, the mass is still inhomogeneous but its gradient is perpendicular to $\beta^\nu$. Since at local equilibrium mass is only a function of temperature, then  $\beta^\nu\nabla_\nu m=0$ can be rewritten as $\beta(\partial m/\partial \beta) u^\nu\nabla_\nu \beta=0$, with $\beta=1/T$. As $\beta(\partial m/\partial \beta)$ is nonzero, it is required that $u^\nu\nabla_\nu \beta=0$, which indicates a purely rotation fluid with the gradient of temperature perpendicular to flow vorticity.

In general, (\ref{requirement2}) is the constraint set to the spin chemical potential $\omega_{\mu\nu}$, with the thermal vorticity a solution with definite physical meaning, and it returns to solution for the constant mass scenario naturally. The optimal ways to determine the spin chemical potential are through the entropy production or the detailed balance. On one hand, when the entropy reaches maximum, the spin chemical potential is constrained to the thermal vorticity $\omega_{\mu\nu}=\frac{1}{2}\partial_{[\mu}\beta_{\nu]}$, and the Killing condition $\partial^{\{\mu}\beta^{\nu\}}=0$ is obtained simultaneously\cite{Becattini:2014yxa, Hattori:2019lfp, Fukushima:2020ucl}. On the other hand, from the kinetic theory, the spin chemical potential is fixed to thermal vorticity by the detailed balance principle, with the Killing condition satisfied \cite{Weickgenannt:2020aaf, Wang:2020pej}. Possible non-equilibrium effects as well as the stability of other solutions will be discussed in further work. 
\section{Conclusion and Outlook}
\label{s5}
In this paper, influence of dynamically generated chiral condensate in spin polarization is discussed in a collisionless kinetic theory framework. Starting from the NJL model, chiral condensate is introduced through the mean field approximation of the auxiliary field, the kinetic equation of the Wigner function is then derived following the standard procedure. Taking the semiclassical expansion, the transport equations for vector component $\mV_\mu$ and axial-vector component $\mA_\mu$ are derived up to $\mathcal{O}(\hbar)$. At classical level, the inhomogeneous dynamical mass provides the force terms in the Vlasov equation, the familiar Vlasov equation can be recovered by taking the equal-time components. Novel phenomenon appears in the first order transport equation, apart from providing the force term, the inhomogeneous condensate also introduces coupled transport between vector component $\mV_\mu$ and axial-vector component $\mA_\mu$. Thus for an unpolarized initial state, spin can get polarized through inhomogeneous mass distribution. It is already well known that collisions convert spin with the orbital angular momentum, however, in this work we find that the mean field condensate can convert OAM with spin as well. 

The conservation of energy momentum tensor and total angular momentum tensor is also proved. With the dynamical generated condensate, the ensemble average of the energy momentum tensor has contribution from both the fermion part and the condensate. The EMT of fermion part alone is not conserved, indicating energy transfers between fermion and the condensate, with the total EMT conserved. The dynamical generated condensate also contributes to the orbital angular momentum, while the angular momentum conservation relation remains the same. The conservation of EMT and AMT are derived from the kinetic equations. 

The stable solution to the spin polarization generated by dynamical condensate is discussed. The normal global equilibrium solution where spin polarization is enslaved by the thermal vorticity is still a stable solution, while the Killing condition is loosened. Possible non-equilibrium effects as well as the stability of other solutions will be discussed in further work. 

\noindent {\bf Acknowledgement:} We thank Dr. Shi Pu, Dr. Shuzhe Shi and Dr. Yu-Chen Liu for helpful discussions. The work is supported by the NSFC grant Nos.12005112 .

\begin{appendix}
\section{Proof of the EMT and AMT conservation}
\label{appen1}
When evaluating the ensemble average of the energy momentum tensor, one need to calculate terms like $\big\langle\hat{\sigma}\bar{\psi}\psi-\hat{\pi}\bar{\psi}i\gamma^5\psi\big\rangle$. The $\sigma$-term is the trace of the following
\begin{eqnarray}
\label{proof1}
&&\langle\sigma(x){\psi}_\alpha(x)\bar\psi_\beta(x)\rangle
\nonumber\\
&=&\lim_{y\rightarrow 0}\left\langle\frac{1}{2}[\sigma(x+\frac{y}{2})+\sigma(x-\frac{y}{2})]{\psi}_\alpha(x+\frac{y}{2})\bar{\psi}_\beta(x-\frac{y}{2})\right\rangle\nonumber\\
&=&\lim_{y\rightarrow 0}\left\langle\sum_{n=0}^{\infty}\frac{1}{(2n)!}\sigma^{(2n)}(\frac{y}{2})^{2n}{\psi}_\alpha(x+\frac{y}{2})\bar{\psi}_\beta(x-\frac{y}{2})\right\rangle\nonumber\\
&=&\lim_{y\rightarrow 0}\int_p\sum_{n=0}^{\infty}\frac{1}{(2n)!}\sigma^{(2n)}(\frac{y}{2})^{2n} e^{ip\cdot y}W_{\alpha\beta}(x,p)\nonumber\\
&=&\lim_{y\rightarrow 0}\int_p\sum_{n=0}^{\infty}\frac{(-1)^{n}}{(2n)!}\sigma^{(2n)}(\frac{1}{2})^{2n}[\partial_p^{2n} e^{ip\cdot y}]W_{\alpha\beta}(x,p)\nonumber\\
&=&-\lim_{y\rightarrow 0}\int_p\sum_{n=0}^{\infty}\frac{(-1)^{n}}{(2n)!}\frac{1}{2^{2n}}\sigma^{(2n)}e^{ip\cdot y}\partial_p^{2n} W_{\alpha\beta}(x,p)\nonumber\\
&=&-\int_p\sum_{n=0}^{\infty}\frac{(-1)^{n}}{(2n)!}\frac{1}{2^{2n}}\sigma^{(2n)}\partial_p^{2n} W_{\alpha\beta}(x,p)\nonumber\\
&=&-\int_p\cos(\hbar \Delta)\sigma(x) W_{\alpha\beta}(x,p).
\end{eqnarray}
Thus we have $
\langle\sigma\bar{\psi}\psi\rangle
=-\int d^4 p\cos(\hbar \Delta)\sigma \mF$ and $
\langle\sigma\bar{\psi}i\gamma^5\psi\rangle
=\int d^4 p\cos(\hbar \Delta)\pi\mP$. 

In order to prove the equivalence of (\ref{EMT_conservation1}) and (\ref{EMT_conservation2}), one needs to have $
\int d^4p p^\nu\sin(\hbar\Delta)\sigma(x) W(x,p)=\frac{1}{2}\sigma(x)\int  d^4p\partial^\nu W(x,p)$, and likewise for the $\pi$-term, 
\begin{eqnarray}
\label{proof2}
&&\int_p p^\nu\sin(\hbar\Delta)\sigma(x) W(x,p)\nonumber\\
&=&\lim_{y\rightarrow0}\int_p e^{ip\cdot y}p^\nu\sin(\hbar\Delta)\sigma(x) W(x,p)\nonumber\\
&=&\lim_{y\rightarrow0}\int_p  e^{ip\cdot y}p^\nu \sum_{n=0}^\infty \frac{(-1)^n}{(2n+1)!}\frac{\partial^{2n+1}\sigma(x)}{2^{2n+1}}\partial_p^{2n+1} W(x,p)\nonumber\\
&=&-\lim_{y\rightarrow0}\int_p  \sum_{n=0}^\infty \frac{(-1)^n}{(2n+1)!}\frac{\partial^{2n+1}\sigma(x)}{2^{2n+1}}[\partial_p^{2n+1} ( e^{ip\cdot y}p^\nu)]W(x,p)\nonumber\\
&=&-\lim_{y\rightarrow0}\int_p  \sum_{n=0}^\infty \frac{(-1)^n}{(2n+1)!}\partial^{2n+1}\sigma(x)\frac{(iy)^{2n+1}}{2^{2n+1}} p^\nu e^{ip\cdot y}W(x,p)\nonumber\\
&&-\lim_{y\rightarrow0}\int_p  \frac{1}{2}\sum_{n=0}^\infty \frac{1}{(2n)!}\partial^{2n+1}\sigma(x)\frac{(y)^{2n}}{2^{2n}}e^{ip\cdot y}W(x,p)\nonumber\\
&=&\lim_{y\rightarrow0}\int_p  \frac{1}{2}\sum_{n=0}^\infty\frac{1}{(2n)!}\partial^{2n}\sigma(x)\frac{(y)^{2n}}{2^{2n}}e^{ip\cdot y}\partial^\nu W(x,p)\nonumber\\
&=& \frac{1}{2}\lim_{y\rightarrow0}\int_p\frac{\sigma(x+\frac{y}{2})+\sigma(x-\frac{y}{2})}{2}e^{ip\cdot y}\partial^\nu W(x,p)\nonumber\\
&=& \frac{1}{2}\sigma(x)\int_p\partial^\nu W(x,p).
\end{eqnarray}
Momentum integral over the last equation in (\ref{mastereq2}) seems to breaks the total angular momentum conservation, here we show that the RHS of (\ref{AMT_conservation}) is exactly vanishing. Consider only the $\sigma$-term,
\begin{eqnarray}
\label{proof3}
&&\int d^4p \sin(\hbar\Delta)\sigma(x) W(x,p)\nonumber\\
&=&\lim_{y\rightarrow0}\int_p e^{ip\cdot y}\sin(\hbar\Delta)\sigma(x) W(x,p)\nonumber\\
&=&\lim_{y\rightarrow0}\int_p  e^{ip\cdot y}\sum_{n=0}^\infty \frac{(-1)^n}{(2n+1)!}\frac{\partial^{2n+1}\sigma(x)}{2^{2n+1}}\partial_p^{2n+1} W(x,p)\nonumber\\
&=&-\lim_{y\rightarrow0}\int_p  \sum_{n=0}^\infty \frac{(-1)^n}{(2n+1)!}\frac{\partial^{2n+1}\sigma(x)}{2^{2n+1}}[\partial_p^{2n+1} ( e^{ip\cdot y})]W(x,p)\nonumber\\
&=&-\lim_{y\rightarrow0}\int_p  \sum_{n=0}^\infty \frac{(-1)^n}{(2n+1)!}\frac{\partial^{2n+1}\sigma(x)}{2^{2n+1}}\Big[(iy)^{2n+1}\Big]e^{ip\cdot y}W(x,p)\nonumber\\
&=&-\lim_{y\rightarrow0}\int_p  \sum_{n=0}^\infty \frac{i}{(2n+1)!}\partial^{2n+1}\sigma(x)(\frac{y}{2})^{2n+1} e^{ip\cdot y}W(x,p)\nonumber\\
&=&-i \lim_{y\rightarrow0}\int_p \frac{\sigma(x+\frac{y}{2})-\sigma(x-\frac{y}{2})}{2}e^{ip\cdot y}W(x,p)\nonumber\\
&=&0,
\end{eqnarray}
the $\pi$-term is likwise.

\end{appendix}

\bibliographystyle{iopart-num}
\bibliography{ref}

\end{document}